\documentclass[12pt]{article}
\usepackage{epsf}
\usepackage{amsmath,amssymb}
\usepackage{latexsym}
\usepackage{graphicx,color}
\usepackage{wrapfig}
\usepackage{latexsym}
\usepackage{graphics}

\usepackage{color}
\usepackage{delarray,color}
\usepackage{fancybox}  
\newcommand {\beq}{\begin{eqnarray}}
\newcommand {\eeq}{\end{eqnarray}}

\newcommand{\be}{\begin{equation}}
\newcommand{\ba}{\begin{eqnarray}}
\newcommand{\ea}{\end{eqnarray}}
\newcommand{\ee}{\end{equation}}

\newcommand{\beqa}{\begin{eqnarray}}
\newcommand{\eeqa}{\end{eqnarray}}
\newcommand{\CR}{\nonumber \\}

\newcommand{\unit}{\hbox to 3.8pt{\hskip1.3pt \vrule height 7.4pt
    width .4pt \hskip.7pt \vrule height 7.85pt width .4pt \kern-2.4pt
    \hrulefill \kern-3pt \raise 3.7pt\hbox{\char'40}}}

\def\matt[#1,#2,#3,#4]{\left(%
\begin{array}{cc} #1 & #2 \\ #3 & #4 \end{array} \right)}

\usepackage{tabularx}

\catcode`\@=11
\@addtoreset{equation}{section}

\catcode`@=12
\relax

\begin{document}
\baselineskip 0.7cm

\begin{titlepage}

\setcounter{page}{0}

\renewcommand{\thefootnote}{\fnsymbol{footnote}}

\begin{flushright}
YITP-09-57\\
\end{flushright}

\vskip 1.35cm

\begin{center}
{\Large \bf 
M5-brane Solution in ABJM Theory and Three-algebra
}

\vskip 1.2cm 

{\normalsize
Seiji Terashima$^1$\footnote{terasima(at)yukawa.kyoto-u.ac.jp} and Futoshi Yagi$^1$\footnote{futoshi(at)yukawa.kyoto-u.ac.jp}
}

\vskip 0.8cm

{ \it
$^1$Yukawa Institute for Theoretical Physics, Kyoto University, Kyoto 606-8502, Japan
}

\end{center}

\vspace{12mm}

\centerline{{\bf Abstract}}

We construct a new classical solution in the ABJM theory
corresponding to M5-branes with a non-zero self-dual three-form flux.
This is an M-theory lift of the D4-brane solution
expressed as a non-commutative plane
in the three dimensional super Yang-Mills theory.
We discuss that our solution is closely related with
the three-algebra.
We show that the corresponding configuration of the M5-brane
satisfies the equations of motion in the single M5-brane action.
We find the agreement between the tension 
of the M5-brane solution in the ABJM action 
and the one computed from the single M5-brane action.

\end{titlepage}
\newpage

\tableofcontents

\section{Introduction}

M-theory is still mysterious although 
its existence was conjectured fifteen years ago \cite{Witten:1995ex}. 
Recently, the action of the multiple M2-branes in M-theory was found by 
Aharony, Bergman, Jafferis and Maldacena (ABJM) \cite{ABJM} 
after the ground-breaking 
works of Bagger and Lambert \cite{BL1,BL2,BL3} and Gustavsson \cite{G}.
This ABJM action is useful to consider the $AdS_4/CFT_3$ duality and 
will be important to understand M-theory.

On the other hand, on the M5-brane there is a self-dual three-form field strength
and the action of multiple M5-branes is also interesting, however,
we know little about it. 
To study it, the ABJM action will be useful because
the bound state of the M5-branes and the M2-branes 
can be described by the M2-brane action, where the M5-branes will be 
represented as "solitons". 
Indeed, the BPS solution corresponding the funnel type bound state of these 
were found in \cite{Terashima}, which can be regarded as a variant of
the famous solution in the BLG action by Basu and Harvey \cite{Basu:2004ed},
and have been studied further in \cite{GRVV, HL}.
(Other solitons in the ABJM action also have been found in \cite{Fujimori}-\cite{Auzzi}.)
This is the M-theory lift of the bound state of the 
D2-branes and the D4-branes which are described as the solution of the
Nahm equation
from the D2-brane point of view or the monopole from the D4-brane point
of view.
The shape of the solution is (fuzzy) $S^3/Z_k$ at a point in the world volume 
of the M2-branes\footnote{
In \cite{Nastase:2009ny} it was shown that we have fluctuations on $S^2$ instead of $S^3/Z_k$
in the perturbative analysis. 
To see the correct fluctuation the full analysis of 
Chern-Simons-matter action with finite $k$ seems to be needed.}
and this fuzzy $S^3/Z_k$ solutions are the ground states of 
the mass deformed ABJM action \cite{GRVV}.
We hope that from these solutions we will find
useful description of the M5-branes although we have not found it.

Other than the funnel like solution, there is a simpler bound state 
of D2-branes and D4-branes, i.e. 
D4-branes with a constant magnetic field or 
infinitely many D2-branes with $[X^1,X^2]=const.$ where
$X^i$ are matrix valued scalar fields representing the position of D2-branes.
This looks like a non-commutative plane.
We expect that there 
is an M-theory lift of this in type IIA string theory.

In this paper, we consider solutions of the equations of motion in the
ABJM action
corresponding to this bound state.\footnote{
In the BLG model, an M5-brane was constructed using 
the Nambu-bracket as the three-algebra \cite{HM, HIMS}.
However, multiple M5-branes have not been obtained
in the BLG model
and the solution is reduced to the D4-brane action with
the Poisson bracket, instead of the *-product commutator,
which should be appeared in the full D4-brane action with flux.
}%
\footnote{For related work, see also \cite{PSST}-\cite{Krishnan:2008zm}.}
These solutions are identified as the bound state of the M5-branes and M2-branes
which becomes the bound state of the D4-branes and D2-branes 
in the well known scaling limit with $k \rightarrow \infty$.
However, the solutions in the ABJM action are non-BPS and non-flat
because of the non-trivial correspondence with the M2-branes in $R^8/Z_k$
and the D2-branes in $R^7$ in the scaling limit.

The solution is given by a perturbative series and 
we show that the existence and uniqueness of the solution.
In order to show the existence of the solution,
the three-algebra structure in the ABJM action \cite{BL4}
and some additional identities are important.
We also find the full solution in the limit where the flux goes to infinity.
We find the agreement between the tension of the M5-brane solution 
in the ABJM action and the one computed from M5-brane world volume action.

It is interesting that
the three-bracket evaluated for the M5-brane solution in the ABJM action
becomes the self-dual three-form flux in the M5-brane point of view.
This can be considered as an M-theory analogue of the fact that 
the two-bracket, i.e. the commutator $[X^1,X^2]$, 
corresponds to the two-form flux $F$ in the D4-brane.
This may indicate that the three-algebra is a part of 
the algebra which describes the multiple M5-brane action.

Instead of the M5-brane, 
we also find that a solution on the ABJM action which becomes
the D2-brane with constant flux in the scaling limit.
In our simple ansatz, it should have a light-like flux 
in the scaling limit, i.e. the bound state with a D2-brane,
fundamental strings and D0-branes. We observed 
that the solution in the ABJM action correctly becomes
the M2-brane winding $S^1$ which is needed to lift type IIA string
theory to M-theory with the momentum in the $S^1$.

The organization of this paper is as follows:
In section 2, we briefly review the ABJM action 
and its reduction to the D2-brane world volume theory.
In section 3, we show a classical solution of 
the equations of motion in the ABJM action
and discuss that it can be interpreted as M5-branes.
In section 4, we consider the corresponding M5-brane
from the single M5-brane action and
show that the corresponding configuration
with a suitable self-dual three-form flux satisfies its equations of motion. 
We also see agreement of the M5-brane tension computed from 
the ABJM action and the one from the M5-brane action.
In section 5, we consider another solution, 
which correspond to M2-brane winding the M-circle 
with momentum in the M-circle.
Section 6 is devoted to conclusion and discussion.
In Appendices, we consider normalization factor 
of gauge coupling constant and scalar fields,
which are needed to estimate the M5-brane tension
including the numerical factor.

\section{A Brief Review of the ABJM Action}

In this section, we briefly review the ABJM action \cite{ABJM},
which describe the M2-branes in the low energy limit. 
This is three-dimensional ${\cal N}=6$ Chern-Simons 
matter theory, whose gauge group is $U(N) \times U(N)$.
We denote the corresponding gauge fields 
as $A^{(1)}$ and $A^{(2)}$ and 
bi-fundamental scalar fields as $Y^A$,
where $A=1,2,3,4$.

The bosonic part of the ABJM action is given by
\begin{eqnarray}
L= \frac{k}{4\pi} \varepsilon^{\mu\nu\rho}
\mathrm{tr} \left( A^{(1)}_{\mu} \partial_{\nu} A^{(1)}_{\lambda} 
+ \frac{2i}{3} A^{(1)}_{\mu} A^{(1)}_{\nu} A^{(1)}_{\lambda} 
- A^{(2)}_{\mu} \partial_{\nu} A^{(2)}_{\lambda} 
- \frac{2i}{3} A^{(2)}_{\mu} A^{(2)}_{\nu} A^{(2)}_{\lambda} \right) \nonumber \\
- \mathrm{tr} \left[ ( D_{\mu} Y_A )^{\dagger} D^{\mu} Y^A \right]
- V_{\rm bos}
\label{ABJM_action}
\end{eqnarray}
where the bosonic potential $V_{\rm bos}$ is given by \cite{ABJM, Benna:2008zy}
\begin{eqnarray}
V_{\rm bos} = 
- \frac{4\pi^2}{3k^2} \mathrm{tr} 
\left[ Y^A Y^{\dagger}_A Y^B Y^{\dagger}_B Y^C Y^{\dagger}_C 
+ Y^{\dagger}_A Y^A Y^{\dagger}_B Y^B Y^{\dagger}_C Y^C \right. 
\qquad\qquad \nonumber \\
 \left. +4 Y^A Y^{\dagger}_B Y^C Y^{\dagger}_A Y^B Y^{\dagger}_C 
- 6 Y^A Y^{\dagger}_B Y^B Y^{\dagger}_A Y^C Y^{\dagger}_C \right].
\label{Potential}
\end{eqnarray}

The moduli space of this theory is $(\mathbb{C}^4/\mathbb{Z}_k)^N/S_N$,
where $\mathbb{Z}_k$ corresponds to the simultaneous
rotation of the phases of scalar fields $Y^A$.
Thus, this model is suggested to describe $N$ M2-branes
probing $\mathbb{C}^4/\mathbb{Z}_k$.

When we take the limit of $k \to \infty$ 
and look at the point 
infinitely far away from the orbifold fixed point
of this $\mathbb{C}^4/\mathbb{Z}_k$,
the geometry can be locally regarded as a cylinder. 
Thus, the ABJM model in this limit is expected to 
describe $N$ D2-branes in type IIA superstring theory.
Indeed, when we give a vacuum expectation value $v$ to one of 
the scalars $Y^i$ and expand around that vacuum,
we obtain the well known D2-brane world volume theory 
in the following limit \cite{Mukhi, ABJM};
\begin{eqnarray}
k,v \to \infty  \,\, \mathrm{with} \qquad \frac{k^2}{32 \pi^2 v^2}
=\frac{1}{4 g_{YM}^2} \,\, \mathrm{fixed}. 
\label{limit1}
\end{eqnarray}
Here the relation between the radius $R$ of the compactified $S^1$
and the string coupling $g_s$ and string scale $l_s$ 
in the type IIA string theory is 
$ R=g_s l_s, \,\,\, l_p=g_s^{ \frac{1}{3} } l_s$
or 
\beq
g_s=\left( \frac{R}{l_p} \right)^{\frac{3}{2}},
\,\,\,\, l_s= \left( \frac{l_p^3}{R} \right)^{\frac{1}{2}},
\label{M_relation}
\eeq
and we have 
\beq
R^2=l_p^3 \frac{8\pi^2v^2}{k^2}
\label{Radius}
\eeq
and the tension of the D2-brane 
is $\tau_{D2}=\frac{1}{g_s l_s^3}=\frac{1}{l_p^3}$
where because $[Y^i]=[v]=[L^{-\frac{1}{2}}]$.
The precise normalization constants appeared above are 
explained in the Appendix A and Appendix B.
Note that the field independent term for the D2-brane action,
i.e. $L_{0}=\tau_{D2} N$, is reproduced in the limit 
if we include $L_{0}=\frac{1}{(2 \pi)^2 l_p^3} tr(1)$ to the ABJM action.
In the following sections, 
we will take $l_p=1$ and $[Y^i]=[v]=[L]$
for convenience.

\section{M5-brane Solution in the ABJM Action}

In this section, we find the solution to the 
equations of motion of ABJM action, which will correspond 
to an M5-brane constructed from infinitely many M2-branes.
Note that a solution for $N$ M5-branes is easily obtained 
by the direct sum of $N$ copies of 
the single M5-brane solution as for the D4-D2 bound state.

It is known that a higher dimensional D-brane with magnetic flux on it 
can be constructed from 
infinitely many lower dimensional D-branes. 
For example, a D4-brane with flux can be regarded as 
infinitely many D2-branes.
Corresponding to this fact, 
the solution of the $N$ D2-brane world volume theory 
corresponding to the D4-brane with flux is found 
in the large $N$ limit.
This solution is given by 
\footnote{The normalization factor of this solution is discussed in
Appendix \ref{NC_param}.}
\begin{eqnarray}
X^1 \sim \hat{x}, \,\,\, X^2 \sim \hat{y}, 
\end{eqnarray}
where $X^i$ is the $N \times N$ matrix valued coordinates of the D2-branes
and $\hat{x}$ and $\hat{y}$ are non-commutative quantities
satisfying
\begin{eqnarray}
[\hat{x},\hat{y}] = i\Theta.
\label{D4}
\end{eqnarray}
The D4-brane is expanded to 4+1 dimensional Minkowski space-time
and extra two directions are described by eigenvalues of $\hat{x}$ and $\hat{y}$.
Similar situation is expected for M-theory;
an M5-brane can be constructed from infinitely many 
M2-branes.

If the ABJM action really describes M2-branes, 
there should be the solution corresponding to an M5-brane.
To be consistent, such a solution 
should reduce to D4-brane solution (\ref{D4})
in the limit of $S^1$ compactification (\ref{limit1}).
Thus, we have already known the M5-brane solution
in the leading order of $v$.
The strategy is to solve 
the equations of motion of ABJM action perturbatively in $1/v$.

In the following, we solve the equations of motion of 
the ABJM action.
We put the following ansatz
\begin{eqnarray}
&&Y^1=Y_1^{\dagger}, \quad Y^2=Y_2^{\dagger}, 
\qquad
\partial_{\mu} Y^1 = \partial_{\mu} Y^2 = 0, 
\nonumber \\  
&&Y^3=Y^4=0, \nonumber \\  
&&A^{(1)}_{\mu} = A^{(2)}_{\mu} = 0, 
\label{ansatz1}
\end{eqnarray}
and solve the equations of motion.
As we will see in later the solution will extend in 
the gauged $U(1)$ direction $Y^i \rightarrow e^{i \theta} Y^i$, thus,
this ansatz will describe a static M5-brane extending in
$(x^0,x^1,x^2)$ and 
\beq
Y^1 = r e^{i \theta}, \;\; Y^2 = r' e^{i \theta}, \;\; Y^3 = Y^4 = 0,
\label{M5_config}
\eeq
where $0 < r < \infty, \; 0 < r' < \infty, \; 0 \le \theta < 2 \pi/k$.
Since in the large $v$ limit the solution will be the 
non-commutative D4-brane solution in the D2-branes, we further impose
\begin{eqnarray}
&&Y^1 = v + \hat{x} + f(\hat{x}, \hat{y}) \nonumber \\
&&Y^2 = \hat{y} \label{ansatz2}
\end{eqnarray}
where $\hat{x}$ and $\hat{y}$ are non-commutative quantity satisfying
(\ref{D4}). Here we regard
$f= {\cal O} (1/v)$ and $\hat{x}, \hat{y}, \Theta ={\cal O}(1)$.
Note that by redefine $\hat{x}$ perturbatively in $1/v$ 
such that $[\hat{x}, Y^2 ]=i\Theta$ we can always choose $Y^2=\hat{y}$.
We also note that the constant and $\hat{x}$ terms 
are special in the ansatz (\ref{ansatz2}), then
we should further impose that $f$ does not contain such terms
in, for example, the Weyl order.

We note that a solution of multiple M5-branes
can be easily obtained from the solution of an M5-brane.
Indeed, 
the following is obviously the solution for $M$ M5-branes:
\beqa
Y^1 &=& 1_M \otimes \left( v + \hat{x} + f(\hat{x}, \hat{y}) \right)
\nonumber \\
Y^2 &=& 1_M \otimes \hat{y},
\eeqa
where $1_M$ is the $M \times M$ unit matrix.
Below, we will consider an M5-brane solution.

As shown in \cite{Terashima}, the bosonic potential
term can be conveniently rewritten as 
\begin{eqnarray}
V_{\rm bos} = 
- \frac{2\pi^2}{3k^2} 
{\rm Tr} \left[ \left( \tilde{Y}_A (\tilde{Y}_B\tilde{Y}_B) 
- (\tilde{Y}_B\tilde{Y}_B) \tilde{Y}_A \right) ^2
-2 \left( \tilde{Y}_A \tilde{Y}_B \tilde{Y}_C 
- \tilde{Y}_C \tilde{Y}_B \tilde{Y}_A \right) ^2
\right] ,
\end{eqnarray}
where $\tilde{Y}_A$ is the $2N \times 2N$ Hermitian matrices
given by
\begin{eqnarray}
\tilde{Y}_A
= \left(
\begin{array}{cc}
0 & Y^A \\
Y_A^{\dagger} & 0
\end{array}
\right) .
\end{eqnarray}
For $Y^3=Y^4=0$,
this potential term reduced to
\begin{eqnarray}
V =  \frac{2\pi^2}{k^2} 
{\rm Tr} \left( [\tilde{Y}^1,(\tilde{Y}^2)^2]^2 + [\tilde{Y}^2,(\tilde{Y}^1)^2]^2 \right)
\label{pot}
\end{eqnarray}
Then we can easily see that 
what we should solve under these ansatz are the following two equations:
\begin{eqnarray}
0 = \frac{\partial V_{\rm bos}}{\partial Y^1} =  \left[ (Y^2)^2, [Y^1, (Y^2)^2] \right] + \left\{ Y^1, [Y^2, [(Y^1)^2 , Y^2] \right\} \label{Y1} \\
0 = \frac{\partial V_{\rm bos}}{\partial Y^2} =  \left[ (Y^1)^2, [Y^2, (Y^1)^2] \right] + \left\{ Y^2, [Y^1, [(Y^2)^2 , Y^1] \right\} \label{Y2}
\end{eqnarray}
We expand $Y_1$ and $Y_2$ with regard to $v$.
Rescaling as 
$\hat{x} = v \tilde{x}$, $\hat{y}=v \tilde{y}, f= v \tilde{f}, \Theta=v^2 \tilde{\Theta}$,
which means 
$Y^1=v(1+\tilde{x}+\tilde{f}), Y^2=v \tilde{y}, [\tilde{x},\tilde{y}]=i\tilde{\Theta}$,
and substituting the ansatz into the equations, the $v$ dependence is 
factored out and we obtain rather
complicated two equations:
\begin{eqnarray}
0 &=& (4\tilde{\Theta}^2 + 4\tilde{\Theta}^2 \tilde{x}) \nonumber \\
&& + \Bigl( - 4 [\tilde{y},[\tilde{y},\tilde{f}]] 
+ 8i\tilde{\Theta} [\tilde{y},\tilde{f}] 
- 4\left\{ \tilde{x},\left[\tilde{y},[\tilde{y},\tilde{f}] \right] \right\} 
+ 4\tilde{\Theta}^2\tilde{f} + 4i\tilde{\Theta}\{ \tilde{x},[\tilde{y},\tilde{f}] \} \nonumber \\
&& \qquad - \{\tilde{x}, \{\tilde{x}, [\tilde{y}, [\tilde{y} , \tilde{f}]]\} \} 
- \{\tilde{y}, [\tilde{y},[ \tilde{y} , [\tilde{y} , \tilde{f}]]\}\} \Bigr) \nonumber \\
&& + \Bigl( -4[\tilde{y},\tilde{f}]^2 - 4\{ \tilde{f},[\tilde{y},[\tilde{y},\tilde{f}]]\} 
- 2 \{ \tilde{x},[\tilde{y},\tilde{f}]^2 \} 
+ 4i\tilde{\Theta} \{ \tilde{f}, [\tilde{y},\tilde{f}]\} \nonumber \\
&& \qquad - \{ \tilde{x},\{\tilde{f},[\tilde{y},[\tilde{y},\tilde{f}]]\}\} 
- \{ \tilde{f}, \{ \tilde{x},[ \tilde{y}, [\tilde{y}, \tilde{f}]]\}\} \Bigr) \nonumber \\
&& + \Bigl( -2\{\tilde{f}, [\tilde{y},\tilde{f}]^2 \} 
- \{ \tilde{f}, \{ \tilde{f}, [\tilde{y}, [\tilde{y}, \tilde{f}]] \} \}  \Bigr) \label{eq1} 
\end{eqnarray}
\begin{eqnarray}
0 &=& (4\tilde{\Theta}^2 \tilde{y}) \nonumber \\
&& + \Bigl( 4 [\tilde{x},[\tilde{y},\tilde{f}]] 
+ 4\left\{ \tilde{x},\left[\tilde{x},[\tilde{y},\tilde{f}] \right] \right\} 
+ 4i\tilde{\Theta}\{ \tilde{y},[\tilde{y},\tilde{f}] \} \nonumber \\
&& \qquad + \{\tilde{y}, \{\tilde{y}, [\tilde{x}, [\tilde{y} , \tilde{f}]]\} \} - \{\tilde{x}, [\tilde{x},[ \tilde{x} , [\tilde{y} , \tilde{f}]]\}\} \Bigr) \nonumber \\
&& + \Bigl( 4[\tilde{f},[\tilde{y},\tilde{f}]] + 4 \{ \tilde{f}, [ \tilde{x}, [\tilde{y},\tilde{f}]]\} + 4 \{ \tilde{x}, [\tilde{f} , [\tilde{y},\tilde{f}]]\} + \{\tilde{x}, \{ \tilde{x},[\tilde{f},[\tilde{y},\tilde{f}]]\}\} \nonumber \\
&& \qquad - 2 \{ \tilde{y}, [\tilde{y},\tilde{f}]^2 \} + \{ \tilde{y}, \{ \tilde{y}, [ \tilde{f}, [\tilde{y},\tilde{f}]]\}\} + \{ \tilde{x}, \{ \tilde{f}, [ \tilde{x}, [\tilde{y} , \tilde{f}]]\}\} 
+ \{ \tilde{f}, \{ \tilde{x},[\tilde{x},[\tilde{y},\tilde{f}]]\}\} \Bigr) \nonumber \\
&& + \Bigl( 4 \{ \tilde{f},[\tilde{f},[\tilde{y},\tilde{f}]]\} + \{ \tilde{x}, \{ \tilde{f},[\tilde{f},[\tilde{y},\tilde{f}]]\}\} + \{ \tilde{f}, \{ \tilde{f}, [\tilde{x}, [ \tilde{y},\tilde{f}]]\}\} 
+ \{\tilde{f}, \{\tilde{x}, [\tilde{f}, [\tilde{y},\tilde{f}] ] \} \} \Bigr) \nonumber \\
&& + \Bigl( \{\tilde{f}, \{\tilde{f} , [ \tilde{f}, [\tilde{y},\tilde{f}]]\}\} \Bigr) \label{eq2}
\end{eqnarray}

\subsection{Existence of the solution}\label{ambig}

In this subsection, we discuss that the 
solution of (\ref{eq1}) and (\ref{eq2}) exist
at least perturbatively.
We also discuss the parameters of the perturbative solution
which we found.

The lowest order terms are $\tilde{\Theta}^2$ and $[\tilde{y},[\tilde{y},\tilde{f}]]$ in (\ref{eq1})
and $[\tilde{x},[\tilde{y},\tilde{f}]]$ in (\ref{eq2}) which are ${\cal O} (1/v^4)$.
From these terms which are linear in $\tilde{f}$,
we can solve the eq. (\ref{eq1}) or the eq. (\ref{eq2})
by choosing $\tilde{f}$ perturbatively in $1/v$,
except terms including
$\tilde{f} \sim \tilde{x} h_1 (\tilde{y})+h_0(\tilde{y})$
for (\ref{eq1}) or 
$\tilde{f} \sim h_2 (\tilde{x})+h_0(\tilde{y})$
for (\ref{eq2}).
Here, one would not expect these two equations are compatible
because these two complicated equations do not seem to be resembled.
However, we can show these are indeed compatible and solutions exist
for them.
The essential identity for ensuring the existence of the solution is
\begin{eqnarray}
\left[ \frac{\partial V_{\rm bos}}{\partial Y^1} , Y^1 \right]
+ \left[ \frac{\partial V_{\rm bos}}{\partial Y^2} , Y^2 \right] =0,
\label{common}
\end{eqnarray}
which can be shown from (\ref{eq1}) and (\ref{eq2}) explicitly.
This equation implies that 
$\tilde{f}$ determined perturbatively in $1/v$
from eq.(\ref{eq1}) 
is same as the one determined from eq.(\ref{eq2}) 
except for the following terms 
$\tilde{f} \sim h_2 (\tilde{x})+\tilde{x} h_1
(\tilde{y})+h_0(\tilde{y})$
because $\left[ \frac{\partial V_{\rm bos}}{\partial Y^1} , Y^1
\right] \sim \left[ {\rm l.h.s. \, of \, eq.(\ref{eq1})} , \hat{x}
\right]$.
Therefore, we conclude that the solutions exist.

From the above discussion, 
we see that the solutions has ``integration constant''
as $\tilde{f} \sim C \tilde{x}+h_0(\tilde{y})$
where $C$ is an arbitrary constant and $h_0$
is an arbitrary function. 
Here we note that 
a redefinition of $\hat{x}=\hat{x}' + h(\hat{y})$,
which is a coordinate transformation,
does not change the commutation relation $[\hat{x}',\hat{y}]=i\Theta$.
This is also regarded as 
a gauge transformation whose gauge parameter depends only on $\hat{y}$ as
\begin{eqnarray}
Y^i \to e^{i\alpha(\hat{y})} Y^i e^{-i\alpha(\hat{y})}.
\end{eqnarray}
Thus we can eliminate the term $h_0$ depending only on
$\hat{y}$ in $Y^1$
without changing (\ref{ansatz2}).
The terms
$\tilde{f} \sim C \tilde{x}+constant$
should be set to zero because the constant and $\hat{x}$ 
terms in $Y^1$ are already extracted in the ansatz (\ref{ansatz2}).
Then, if we make a redefinition of $\hat{x}=\alpha \hat{x}' + \beta$,
where $\alpha, \beta$ are constants,
we have $Y^1=v'(\beta)  + \gamma_1(\alpha,\beta) \hat{x}'+
\dots$ and $[\hat{x}',\hat{y}]= i\Theta'(\alpha,\beta)$. 
Taking $\gamma_1(\alpha,\beta)=1$ by choosing $\alpha$ appropriately,
we still have one parameter $\beta$, which can be used 
to fix $\Theta'(\alpha,\beta)=1$. 
Therefore, the two parameters $v,\Theta$ in the ansatz (\ref{ansatz2})
are not independent and we conclude that
the solutions exist at least perturbatively 
with one physical parameter corresponding
to the flux of the corresponding D4-brane solution.
Indeed, $v$ is the parameter where we expand $Y^1$ around
and then is not physical meaning itself.

The identity (\ref{common}) is essential for solving the 
equations. An analogue of this for the D2-branes 
where $V_{\rm bos} \sim [Y^i,Y^j]^2$ will be
$[Y^i, [Y^j,[Y^i,Y^j]]]=0$, which follows from 
the Jacobi identity. 
For the ABJM action,
the three-algebra structure will be important.
The identity (\ref{common}) can be shown to be derived from
the fundamental identity of the three-algebra with some
identities including the two-algebra and three-algebra.
We will show it below.
Following \cite{BL4} we define a three-bracket
\beq
[\tilde{Y}_A,\tilde{Y}_B,\tilde{Y}_C] \equiv 
\tilde{Y}_A \tilde{Y}_B \tilde{Y}_C -\tilde{Y}_C \tilde{Y}_B \tilde{Y}_A,
\label{3bracket}
\eeq
and an inner product $(\tilde{Y}_A, \tilde{Y_B}) 
\equiv {\rm Tr} (\tilde{Y}_A \tilde{Y_B})$ 
for $2N \times 2N$ Hermitian matrices $\tilde{Y}$,
then the fundamental identity of the three-algebra becomes
\beq
0=[[A,B,C],D,E]-[[A,D,E],B,C]-[A,B,[C,D,E]]-[A,[B,E,D],C].
\eeq
Moreover $[A,B,C]=-[C, B,A]$ and $([A,B,C],D)=-(A,[B,C,D])$
are satisfied identically.
We can also show that
\begin{eqnarray}
0 &=& [A,[A,B],[A,A,B]]+[[A,B],A,[A,A,B]] \CR
0 &=& [A,[A,B],[A,A,B]]+2 [A,B,[A,A,[A,B]]] \CR
&&-[A,A,[A,B,[A,B]]] - [B,A,[A,A,[A,B]]].
\end{eqnarray}
The l.h.s. of (\ref{common}) is proportional 
to $2 [A,[A,B],[A,A,B]] + 2 [A,B,[A,A,[A,B]]]
- [A,A,[A,B,[A,B]]] - [B,A,[A,A,[A,B]]] + [[A,B],A,[A,A,B]]$
which identically vanishes from the above identities.

\subsection{Perturbative solution}

In order to compute $Y^1$,
it is convenient to use Weyl ordered product 
$\{ \hat{x}^n \hat{y}^m \}_W$,
which is defined as
\begin{eqnarray}
\exp (\alpha x + \beta y) = \sum_{n,m} \frac{1}{n!m!} \alpha^n \beta^m 
\{ \hat{x}^n \hat{y}^m \}_W
\end{eqnarray}
When we rewrite usual product of Weyl ordered products
into Weyl order, we can use the formula of star product.
When we write the function replacing 
the product of $x$ and $y$ of the function $f(x,y)$ and $g(x,y)$
into the Weyl ordered product of $\hat{x}$ and $\hat{y}$ 
as $f(\hat{x},\hat{y})_W$, $g(\hat{x},\hat{y})_W$,
\begin{eqnarray}
f(\hat{x},\hat{y})_W \,\, g(\hat{x},\hat{y})_W = h(\hat{x},\hat{y})_W,
\end{eqnarray}
where 
\begin{eqnarray}
h(x,y) = f*g \, (x,y).
\end{eqnarray}
In this way the function $h(x,y)$ is given by the 
star product of $f(x,y)$ and $g(x,y)$.
Corresponding to the commutation relation
$[\hat{x}, \hat{y}] = i\Theta$,
the star product is given as
\begin{eqnarray}
f*g(x,y) 
&=& \sum_{n=0}^{\infty} \sum_{k=0}^n \frac{(-1)^{n-k}}{(n-k)! \, k!}
\left( \frac{i\Theta}{2}\right)^n \frac{\partial^n f(x,y)}{\partial x^k \partial y^{n-k}} 
\frac{\partial^n g(x,y)}{\partial x^{n-k} \partial y^k} \nonumber \\
&=& f g + \frac{i\Theta}{2} \left( 
\frac{\partial f}{\partial x} \frac{\partial g}{\partial y} 
- \frac{\partial f}{\partial y} \frac{\partial g}{\partial x} 
\right) + \cdots .\label{def_star}
\end{eqnarray}

Following the ansatz (\ref{ansatz2}) and the convention
discussed in subsection \ref{ambig},
we expand $Y^1$ with regard to $v$ as 
\begin{eqnarray}
Y^1 = v + \hat{x} + \sum_{n=2}^{\infty} v^{1-n} 
\sum_{k=0}^{[(n-2)/4]} (i\Theta)^{2k} 
\sum_{m=1}^{n-4k} a_{n,m,k} \hat{x}^m \hat{y}^{n-m-4k}
\end{eqnarray}
where we write $\{\hat{x}^n \hat{y}^m\}_W$ as $\hat{x}^n \hat{y}^m$ for simplicity.
Substituting this into (\ref{Y1}) and (\ref{Y2}),
using the formula (\ref{def_star}), 
and determining the coefficients perturbatively, 
the function $Y^1$ is written using the
Weyl ordered product as follows.
\begin{eqnarray}
Y^1 &=& v + \hat{x} - \frac{\hat{x}^2}{2v} + \left( \frac{1}{2}\hat{x}^3 
-\frac{1}{2} \hat{x}\hat{y}^2 \right) \frac{1}{v^2} 
+ \left( -\frac{5}{8} \hat{x}^4 + \hat{x}^2\hat{y}^2 \right) \frac{1}{v^3} \nonumber \\
&&+ \left( \frac{7}{8} \hat{x}^5 - \frac{23}{12} \hat{x}^3\hat{y}^2 
+ \frac{3}{8} \hat{x}\hat{y}^4 \right) \frac{1}{v^4} 
+ \left( -\frac{21}{16} \hat{x}^6 + \frac{11}{3} \hat{x}^4 \hat{y}^2 
- \frac{3}{2} \hat{x}^2 \hat{y}^4 - \frac{3}{4}\Theta^2 \hat{x}^2 \right) 
\frac{1}{v^5} \nonumber \\
&& + \left( \frac{33}{16} \hat{x}^7 - \frac{563}{80}\hat{x}^5\hat{y}^2
+ \frac{217}{48}\hat{x}^3\hat{y}^4 - \frac{5}{16}\hat{x}\hat{y}^6
+ \frac{13}{8} \Theta^2 \hat{x}^3 \right) \frac{1}{v^6} \nonumber \\
&&+\left( -\frac{429}{128} \hat{x}^8 + \frac{1627}{120}\hat{x}^6 \hat{y}^2
- \frac{145}{12}\hat{x}^4\hat{y}^4 + 2\hat{x}^2\hat{y}^6 
- \Theta^2 \left( \frac{87}{8}\hat{x}^4 - \frac{5}{2} \hat{x}^2\hat{y}^2 \right) \right)  \frac{1}{v^7} \nonumber \\
&& + \left( \frac{715}{128}\hat{x}^9 - \frac{88069}{3360}\hat{x}^7\hat{y}^2
+ \frac{29003}{960}\hat{x}^5 \hat{y}^4 - \frac{273}{32}\hat{x}^3\hat{y}^6
+\frac{35}{128}\hat{x}\hat{y}^8 \right. \nonumber \\
&& \qquad - \left. \Theta^2 \left( - \frac{1291}{40}\hat{x}^5 
+ \frac{143}{8} \hat{x}^3 \hat{y}^2 
+ 2 \hat{x}\hat{y}^4 \right)
\right) \frac{1}{v^8} \nonumber \\
&&+ \left( - \frac{2431}{256} \hat{x}^{10} 
+ \frac{1423}{28}\hat{x}^8 \hat{y}^2 - \frac{52129}{720}\hat{x}^6 \hat{y}^4 
+ 30 \hat{x}^4 \hat{y}^6 - \frac{5}{2} \hat{x}^2 \hat{y}^8 \right. \nonumber \\
&& \left. \qquad - \Theta^2 \left( \frac{8551}{96} \hat{x}^6 - \frac{2005}{24} \hat{x}^4 \hat{y}^2
- \frac{11}{4} \hat{x}^2 \hat{y}^4 \right) - \frac{319}{16} \Theta^4 \hat{x}^2 \right) \frac{1}{v^9}
+ \cdots .
\label{Y1_sol2}
\end{eqnarray}

\subsection{All order solution in the limit of $\Theta \to 0$}

In this subsection, we show that in the limit of $\Theta \to 0$,
we can actually solve the equations of motions
(\ref{Y1}) and (\ref{Y2}) exactly.

Commutator and anti-commutator of star product (\ref{def_star}) can be
approximated at the leading order of $\Theta$ as
\begin{eqnarray}
[f,g]_* &\sim& i\Theta \left( 
\frac{\partial f}{\partial y} \frac{\partial g}{\partial x} 
- \frac{\partial f}{\partial x} \frac{\partial g}{\partial y} 
\right) , \label{app_1} \\
\{ f,g \} _* &\sim& 2fg . \label{app_2}
\end{eqnarray}
When we rewrite (\ref{Y1}) and (\ref{Y2}) using star product 
we obtain
\begin{eqnarray}
&&0 = \left( (Y^1)^2 + y^2 \right) \frac{\partial^2 Y^1}{\partial x^2} 
 + Y^1 \left( \frac{\partial Y^1}{\partial x} \right) ^2 \nonumber \\
&&0= y \left( \frac{\partial Y^1}{\partial x} \right) ^2
 + \left( (Y^1)^2  + y^2 \right) 
   \frac{\partial Y^1}{\partial x} \frac{\partial^2 Y^1}{\partial x\partial y}
 - \left( (Y^1)^2  + y^2 \right) 
   \frac{\partial^2 Y^1}{\partial x^2} \frac{\partial Y^1}{\partial y}
\end{eqnarray}
We find that these two equations are always satisfied simultaneously
when $Y^1$ satisfies 
\begin{eqnarray}
\frac{\partial Y^1}{\partial x} = \frac{v}{\sqrt{(Y^1)^2+y^2}},
\label{exact}
\end{eqnarray}
which give the exact solution in the limit of $\Theta \to 0$.
The solution to this differential equation is obtained 
if we solve the following with regard to $Y^1$
\begin{eqnarray}
vx = \frac{1}{2} Y^1 \sqrt{(Y^1)^2 + y^2} 
    + \frac{1}{2} y^2 \ln \left( Y^1 + \sqrt{(Y^1)^2 +y^2} \right) + g(y) .
\end{eqnarray}
where $g(y)$ is an arbitrary function of $y$.
Because we required $Y^1 |_{x=0} =v$, 
we find $g(y)=-\frac{1}{2} v \sqrt{v^2 + y^2} 
    - \frac{1}{2} y^2 \ln \left( v + \sqrt{v^2 +y^2} \right)$.
Indeed, when we use the perturbative solution (\ref{Y1_sol2}), 
we show
\begin{eqnarray}
\frac{\partial Y^1}{\partial x}   \sqrt{(Y^1)^2+y^2} 
&=& v \left[ 1 - \frac{3}{4} \frac{\Theta^2}{v^4} 
- \frac{3x}{2}\frac{\Theta^2}{v^5}
+ \frac{15x^2}{2}\frac{\Theta^2}{v^6} 
- \left( 29x^3-\frac{17xy^2}{4} \right) \frac{\Theta^2}{v^7} \nonumber \right. \\
&& \qquad 
- (- 97 x^4 + 39 x^2 y^2 + 2 y^4 ) \frac{\Theta^2}{v^8} \nonumber \\
&& \qquad \left. + \left( \frac{67}{16} x y^4 - \frac{653}{3} x^3 y^2 
+ \frac{319}{8} \Theta^2 x + \frac{1478}{5} x^5 \right) \frac{\Theta^2}{v^9}
+ \cdots \right] , \nonumber \\
\end{eqnarray}
which is consistent to the fact that 
(\ref{exact}) is satisfied in the limit of $\Theta\to 0$.
Note that the solution is singular at
$y=0$ and $Y^1=0$, which means $x=-1/2$.
This singularity would be resolved 
by introducing some-branes on which M5-brane ends or slightly modifying
the singularity. 

Of course, we can express the solution without referring 
to the $\hat{x}, \hat{y}$.
The equation of motion becomes 
\beqa
0&=&((Y^1)^2+(Y^2)^2)[Y^2,[Y^1,Y^2]_P ]_P+Y^1 ([Y^1,Y^2]_P)^2, \CR
0&=&((Y^1)^2+(Y^2)^2)[Y^1,[Y^1,Y^2]_P ]_P+Y^2 ([Y^1,Y^2]_P)^2,
\eeqa
where $[Y^1,Y^2]_P$ is the Poisson bracket.
The solution is given by
\beq
[Y^1,Y^2]_P=\frac{C}{\sqrt{(Y^1)^2+(Y^2)^2}},
\eeq
where $C$ is a constant and identified as $C=i\Theta v$ for the explicit 
realization the Poisson bracket.

\subsection{Gauged $U(1)$ direction}

In this subsection, 
we discuss the configuration of the M5-brane
corresponding to the solution found previously.
In order to describe the configuration of the
M5-brane, it is convenient to introduce coordinates
$r,r',\theta,\theta'$, which are related to 
the complex coordinates $Y^1$ and $Y^2$ as
\begin{eqnarray}
Y^1 = r e^{i (\theta+\theta')}, \qquad
Y^2 = r' e^{i (\theta-\theta')}.
\label{param}
\end{eqnarray}

The solution constructed in the above discussion 
seems to expand in $1+4$ dimensional space-time; 
1+2 dimensional space-time which the original M2-branes 
fill and 2 dimensional space which correspond to 
$r$ and $r'$ because $Y^1$ and $Y^2$ are Hermite 
in our solution.
However, such a $1+4$ dimensional object is not
expected in the M-theory.
Since our solution reduces to $D4$-brane solution
in the scaling limit  (\ref{limit1}) by construction, 
it should be interpreted as an M5-brane
extending to the gauged $U(1)$ direction $\theta$,
which reduces to compact $S^1$ direction in the scaling limit.
We expect that we can see this direction
by taking account of the effect of the monopole operators.

As commented in the introduction,
similar situation appears also in \cite{Nastase:2009ny},
where fuzzy 2-sphere appears rather than 3-sphere,
in the large $k$ limit where the perturbative analysis 
is reliable and this is actually the IIA limit.

In the following, we assume that our solution 
describe a static M5-brane extending to the direction
of (\ref{M5_config}). 
The configuration of the M2-branes and the M5-brane 
are summarized in Table \ref{table_M5}.
The induced metric on the M5-brane is given by
\beq
ds^2 = ds_{(3)}{}^2 + dr^2 + dr'^2 + (r^2+r'^2) d\theta^2,
\label{metric}
\eeq
where $ds_{(3)}{}^2$ is the flat metric of the 1+2 dimensional Minkowski space-time.
\begin{table}
\centering
\begin{tabular}{|c|c|c|c|c|c|c|c|c|c|c|c}
\hline
& 0 & 1 & 2 & 3 ($r$) & 4 ($r'$) & 
5 ($\theta$) & 6 ($\theta'$) & 7,8 ($Y^3$)& 9,10 ($Y^4$)\\
\hline
M2-branes & --- & --- & --- & $\bullet$ & $\bullet$ & $\bullet$ 
& $\bullet$ &$\bullet \quad \bullet$ & $\bullet \quad \bullet$ \\
M5-brane & --- & --- & --- & --- & --- & --- 
& $\bullet$ & $\bullet \quad \bullet$ & $\bullet \quad \bullet$ \\
\hline
\end{tabular}
\caption{Configuration of the M2- and M5-branes}
\label{table_M5}
\end{table}
The interpretation that our solution corresponds to an M5-brane
wrapping the gauged $U(1)$ direction $\theta$ 
is justified by calculation of the tension in the next subsection.

\subsection{Tension of the M5-brane solution}

In the following, 
we calculate the tension of the M5-brane 
from our solution.
Approximating the bosonic potential term (\ref{pot})
as the leading order of $\Theta$
using (\ref{app_1}) and (\ref{app_2}), we obtain
\begin{eqnarray}
V_{\rm bos} = \frac{16 \pi^2 \Theta^2}{k^2}  
{\rm Tr} \left[ \left( (Y^1)^2 + y^2 \right) 
\left( \frac{\partial Y^1}{\partial x} \right) ^2 
 \right] 
= \frac{16 \pi^2 \Theta^2 v^2}{k^2}
{\rm Tr} {\bf 1}, \label{pot_const}
\end{eqnarray}
where we used (\ref{exact}).
Using the star product, 
the trace can be replaced as a usual integral  
\begin{eqnarray}
{\rm Tr} \to \int \frac{dxdy}{2 \pi \Theta}.
\label{replace}
\end{eqnarray}
Now we rewrite this in the spacetime coordinate
using the solution (\ref{exact}) in order to interpret it 
as M5-brane,
\begin{eqnarray}
V_{\rm bos} = \frac{8 \pi \Theta v}{k^2}  \int d r d r' \sqrt{r^2+{r'}^2},
\label{V2}
\end{eqnarray}
where $r$ and $r'$ represent $|Y^1|$ and $|Y^2|$, respectively.
Inserting $1=\frac{k}{2 \pi} \int^{\frac{2 \pi}{k}}_{0} d \theta$,
we have 
\begin{eqnarray}
V_{\rm bos} = \frac{\Theta v}{4k} \int d r d r' d \theta \sqrt{r^2+{r'}^2}.
\label{V_TM5}
\end{eqnarray}
Then, by adding the constant term $L_0 = \frac{1}{(2\pi)^2 l_p{}^3} {\rm Tr} {\bf 1}$
to the ABJM action,
the action is evaluated as
\begin{eqnarray}
S = T_{\rm M5} \int d x^0 d x^1 d x^2 d (2\sqrt{2} \pi r) d (2\sqrt{2} \pi r') d \theta 
\sqrt{(2\sqrt{2} \pi r)^2+(2\sqrt{2}\pi r')^2}.
\label{V_TM5p}
\end{eqnarray}
where 
\begin{eqnarray}
T_{\rm M5} = \frac{k}{2\sqrt{2} (2\pi)^7 \Theta v }
+ \frac{\sqrt{2}\Theta v}{(2\pi)^3 k}.
\label{M5_tension}
\end{eqnarray}
The factor $2\sqrt{2} \pi$ is inserted because
$2\sqrt{2}\pi r$ and $2\sqrt{2}\pi r'$
instead of $r$ and $r'$ represent the spacetime length
as explained in Appendix \ref{NC_param}.

Here $T_{M5}$ can be interpreted as the tension of the M5-brane,
which is a constant because r.h.s. of (\ref{V_TM5p}) is indeed
proportional to the volume factor of the corresponding M5-brane
and the period of $\theta$ is $2\pi/k$.
This appearance of the volume factor is evidence that 
our solution indeed corresponds to an M5-brane wrapping 
the gauged $U(1)$ direction $\theta$.
Because the ABJM action is the low energy action, 
the tension (\ref{M5_tension}) will be modified by the 
higher order terms to include ${\cal O}(\Theta^3)$ terms.

\section{Viewpoint from the M5-brane Action}

In this section, we discuss the solution 
found in the previous section in terms of the corresponding M5-brane
by using the single M5-brane action \cite{Perry, Schwarz, PST}.

\subsection{Flux on the M5-brane}

In addition to scalar fields which correspond to the configuration of the M5-brane,
the non-linearly self-dual three-form field strength
\begin{eqnarray}
F_{ijk} = \partial_{i}A_{jk} + \partial_{j}A_{ki}
 +\partial_{k}A_{ij},
\end{eqnarray}
also exists on the M5-brane.
In the following, we use the indices of the space-time coordinates written in Table \ref{table_M5}.
The three-form field strength on the M5-brane 
is related to the dual gauge field strength $f^{ijk}$
 on the D4-brane as  
\begin{eqnarray}
i F_{ijk} = 2 \pi l_s{}^2 f_{ijk} \qquad (i,j,k \neq 5)
\label{F_f}
\end{eqnarray}
in the scaling limit, where "5" becomes the compact $S^1$ direction.
The normalization of the field strength 
$\tilde{f}_{ij}=\frac{1}{6} 
\epsilon_{ijklm} f^{klm}$ on the D4 brane 
is fixed by writing the DBI action as (\ref{DBI}) in Appendix \ref{DBI_SYM}.%
\footnote{In Appendix \ref{DBI_SYM}, we denote the field strength of 
the D4-brane as $F$ instead of $\tilde{f}$.}
It is known that on the D4-brane this flux is equivalent to 
a constant NS-NS $B$ field.
In the large $B$ limit, the non-commutative parameter $\Theta$ is given as 
\begin{eqnarray}
\Theta^{\mu\nu} = \frac{1}{8 \pi^2} \left(\frac{1}{B}\right)^{\mu\nu}
 = \frac{1}{8 \pi^2} \left(\frac{1}{\tilde{f}}\right)^{\mu\nu},
\label{B_theta}
\end{eqnarray}
whose normalization is explained in Appendix \ref{NC_param}.

The index of the space-time coordinate is shown in the Table 1.
In the scaling limit, our solution should reduces to a D4-brane solution,
which is accompanied by a non-zero constant flux 
$\tilde{f}_{34} = f_{012} \neq 0$,
in order to have a constant non-commutative parameter
\beq
\Theta^{34}=\frac{1}{8\pi^2 \tilde{f}_{34}}= 
\frac{1}{8\pi^2 f_{012}}.
\eeq
A candidate of the three-form flux on the M5-brane to reproduce this flux in the scaling limit
is constant flux $F_{012}$.
This constant flux is preferable also from the constant tension (\ref{V_TM5p}).
 $F_{012}$ is related with $F_{345}$ 
via the non-linear self-duality condition\footnote{
In this section, $r$ and $r'$ are taken as $2 \sqrt{2} \pi r$
and $2 \sqrt{2} \pi r'$ in the previous sections, respectively.}
\begin{eqnarray}
F_{012} = - \frac{\tilde{F}_{012}}{\sqrt{1-\tilde{F}_{012}{}^2}}
= - \frac{F_{345}/\sqrt{r^2+r'^2}}{\sqrt{1-F_{345}{}^2/(r^2+r'^2)}}
\label{non_linear}
\end{eqnarray}
or equivalently,
\begin{eqnarray}
F_{345} = - \frac{F_{012}\sqrt{r^2+r'^2}}{\sqrt{1+F_{012}{}^2}}
\label{non_linear2}
\end{eqnarray}
where $\tilde{F}_{ijk}$ is the dual field strength
$$\tilde{F}^{lmn} = \frac{1}{6\sqrt{-g}} \varepsilon^{lmnpqr} F_{pqr}.$$
We assume that components of the flux other than $F_{012}$ and $F_{345}$ vanish.
The explicit expression for the flux which is expected in our M5-brane configuration is
\beq
F_{012}= - \sqrt{-g} \tilde{F}^{345} =
- \frac{E}{\sqrt{1-E^2}}, \;\;
F_{345}= \sqrt{-g} \tilde{F}^{012} =
- E \sqrt{r^2+r'^2},
\label{flux}
\eeq
where $\tilde{F}_{012} \equiv E$ is constant.

\subsection{M5-brane solution}

In the following, we discuss that M5-brane with the configuration (\ref{M5_config})
with the flux (\ref{flux}) satisfies the equations of motion 
obtained from the single M5-brane action \cite{PST}
\begin{eqnarray}
S= \frac{1}{(2 \pi)^5} \int d^6 x \left[ 
\sqrt{-g} \frac{1}{4(\partial a)^2}
\partial_m a(x) \tilde{F}^{lmn} F_{nlp} \partial^p a(x) 
+ \sqrt{- \det (g_{mn}+i\tilde{F}_{mn})}  
\right]
\label{M5_action}
\end{eqnarray}
where $\tilde{F}_{mn}$ is defined as
$$\tilde{F}_{mn} 
= \frac{1}{\sqrt{(\partial a)^2}} \tilde{F}_{mnl} \partial^l a(x)$$
and the field $a(x)$ is an auxiliary field, which can be eliminated
by a gauge transformation.
Although we can check it directly
under the gauge fixing condition $\partial_{\mu} a = \delta_{\mu}^5$,
we will use another condition, which enable us 
to check it easier, in the following.

As the directions $0,1,2$ are flat 1+2 dimensional Minkowski space and
the flux $F_{012}$ is constant, 
we can compactify one of the space directions, say direction 2.%
\footnote{This compactification direction is different from that 
of the reduction from M2 to D2, but to F1. }
This correspond to the gauge fixing $\partial_{\mu} a = \delta_{\mu}^2$ 
but the non-linear relation (\ref{non_linear}) also holds for this case.
The M5-brane becomes a D4-brane and
the flux $F_{012}$ on the M5-brane is reduced
to the electric flux $\tilde{f}_{01} = i\tilde{F}_{012}/(2 \pi l_s^2) $ 
on the D4-brane by this compactification.
It is sufficient to check that this D4-brane configuration 
with a constant flux $\tilde{f}_{01}$ 
is a solution of the DBI action
\begin{eqnarray}
S = \frac{1}{(2\pi)^4 g_s l_s{}^5} \int d^5 \sigma \sqrt{ \det (g_{ab}+\tilde{f}_{ab}) }.
\end{eqnarray}

The D4-brane is extended to the direction 0 and 1,
which are flat Minkowski spacetime,
as can be seen from Table \ref{table_M5},
and the only non-vanishing component of the flux $\tilde{F}_{01}$
is constant.
Thus, if the embedding of the remaining three dimensions
into spacetime spanned by $r$, $r'$, $\theta$, $\theta'$
is the solution of the Nambu-Goto action,
the total configuration becomes the solution of the 
DBI action.
Indeed, we can explicitly check that the configuration $\theta'=0$
with the constant flux $\tilde{F}_{01}$ 
also extremizes the Nambu-Goto action.

Thus, we showed that the configuration of the M5-brane (\ref{M5_config})
with constant flux (\ref{flux}) satisfy the equations of motion.

\subsection{Tension from M5-brane action}

In this section, we calculate the tension 
of the M5-brane using the single M5-brane action
and relate it with the parameters $k,v,\Theta$
appeared in our classical solution of the ABJM model
in the limit of $\Theta \to 0$.
This will give a consistency check by comparing it 
with the tension obtained from our classical solution in the ABJM model.

The M5-brane tension with the induced metric (\ref{metric})
and with the constant flux (\ref{flux})
is obtained from the single M5-brane action (\ref{M5_action}).
Since the first term of the action is topological,
we assume that it does not correspond the ABJM action.
Thus, we estimate only the contribution from the second term.
The result is
\beq
S=T_{\rm M5} \int dx^6 \sqrt{g}
\eeq
where
\begin{eqnarray}
T_{\rm M5} = \frac{1}{(2\pi)^5}
\sqrt{1-\tilde{F}_{34}{}^2}
= \frac{1}{(2\pi)^5} \sqrt{1-F_{012}{}^2}.
\label{TM5_1}
\end{eqnarray}

In the following discussion, we rewrite this with
the parameters $\Theta$, $k$, and $v$ instead of $\tilde{F}_{34}$
in the limit of $\Theta \to 0$.
From (\ref{F_f}) and (\ref{B_theta}) together with (\ref{M_relation})
and (\ref{Radius}),
we obtain a relation between three-form flux and the parameters $\Theta$, $k$, $v$ as
\begin{eqnarray}
iF_{012} = \frac{k}{8\sqrt{2} \pi^2 v \Theta} \label{F_param}
\end{eqnarray}
Substituting (\ref{F_param}) into (\ref{TM5_1})
and taking the limit $\Theta \to 0$,
we obtain 
\begin{eqnarray}
T_{\rm M5} = \frac{k}{2\sqrt{2} (2\pi)^7 \Theta v }
+ \frac{\sqrt{2}\Theta v}{(2\pi)^3 k}
+ {\cal O}(\Theta^3).
\end{eqnarray}
This exactly matches to (\ref{M5_tension})
including the numerical factor.

\subsection{Three-algebra and three-form flux}

In this subsection, we discuss an interesting relation
between the 3-algebra and the flux in our solution.
As discussed in \cite{BL4}, the ABJM bosonic potential 
can be written in terms of the three-bracket as
\beq
V_{\rm bos} = \frac{1}{3} {\rm Tr} (\Upsilon^{CD}_B , \Upsilon^B_{CD})
\eeq 
where 
\beq
\Upsilon^{CD}_B = [Y^C, Y^D; Y_B] 
- \frac{1}{2} \delta_B^C [Y^E, Y^D; Y_E] 
- \frac{1}{2} \delta_B^D [Y^E, Y^C; Y_E].
\eeq
The indices run the complex coordinate as $B,C,D=1,2,3,4$.
Here, the three bracket introduced here is related to the
three bracket defined in (\ref{3bracket}) as
\begin{eqnarray}
[\tilde{Y}^A, \tilde{Y}^B, \tilde{Y}^C] 
= \left(
\begin{array}{cc}
0&[Y^A,Y^C;Y_B] \\ 
{}[Y_A,Y_C;Y^B] & 0
\end{array}
\right),
\end{eqnarray}
where 
\begin{eqnarray}
\tilde{Y}^A = 
 \left(
\begin{array}{cc}
0&Y^A \\ 
Y^\dagger_A & 0
\end{array}
\right).
\end{eqnarray}
Substituting our solution to $\Upsilon^{CD}_B$,
we find 
\beq
\Upsilon^{12}_{1} = \frac{v Y^1}{\sqrt{(Y^1)^2+(Y^2)^2}}, \quad
\Upsilon^{12}_{2} = \frac{v Y^2}{\sqrt{(Y^1)^2+(Y^2)^2}}
\eeq
while other components vanish.
Although we cannot see the gauged $U(1)$ direction $\theta$
explicitly, we make up for them as
\beq
\Upsilon^{12}_{1} = \frac{v re^{i\theta}}{\sqrt{r^2+r'^2}}, \quad
\Upsilon^{12}_{2} = \frac{v r'e^{i\theta}}{\sqrt{r^2+r'^2}}
\eeq
in order that $\Upsilon^{CD}_B$ acts correctly under
the $U(1)$ gauge transformation.
Among the complex directions $1,\bar{1},2,\bar{2}$,
the directions which the M5-brane extends are
$r,r',\theta$.
Calculating the component of $\Upsilon^{CD}_B$ for 
these directions by changing the complex coordinates 
into real coordinates, we obtain
\beq
\Upsilon^{rr'}_{\theta} = iv \sqrt{r^2+r'^2}.
\label{Upsilon}
\eeq

On the other hand, the independent non-vanishing component
of the self-dual three-form flux in our solution is only 
\beq
F^{34}{}_{5} = F_{345} = -E\sqrt{r^2+r'^2}
\label{comp_flux}
\eeq
as in (\ref{flux}).
Comparing (\ref{Upsilon}) and (\ref{comp_flux}),
we find that they match up to a numerical factor.
This indicates that the three-bracket plays a significant role
for an action of the multiple M5-branes.

Note that 
the Nambu bracket on $\{ r,r',\theta \}$
is also proportional to the three-bracket
because  
the three-form flux is proportional to the volume form 
of the induced metric on the M5-brane.

\section{Another Solution in the ABJM Action}

We have studied the solution in the ABJM action
from the solution in the D2-brane action correspond 
to the D4-D2 bound state.
In this section, 
instead of the D4-D2 bound state,
we will consider the solution in the ABJM action
correspond to the D0-D2 bound state,
which is lifted to the M0-M2 bound state,
where we call the Kaluza-Klein momentum 
along the M-circle as the M0-brane.

The D0-D2 bound state is represented by the 
magnetic flux in the D2-branes. 
We will take a D2-brane, i.e. an M2-brane,
thus, we will solve the equations of motion of
the ABJM action with the $U(1) \times U(1)$ gauge group.
We change the basis of the gauge field as
\begin{eqnarray}
A_{\mu} = A_{\mu}^{(1)} + A_{\mu}^{(2)} , \quad 
B_{\mu} = A_{\mu}^{(1)} - A_{\mu}^{(2)}
\end{eqnarray}
The matter fields do not couple to $A_{\mu}$.
We write the field strength as
\begin{eqnarray}
&& F^{\mu\nu} = \partial^{\mu} A^{\nu} - \partial^{\nu} A^{\mu} \\
&& F_B^{\mu\nu} = \partial^{\mu} B^{\nu} - \partial^{\nu} B^{\mu}
\end{eqnarray}
Using these, the Lagrangian can be rewritten as
\begin{eqnarray}
L=\frac{k}{8 \pi} \varepsilon^{\mu\nu\rho} B_{\mu} F_{\nu\rho}
 - ( \partial_{\mu} Y_A + i B_{\mu} Y_A  )^{\dagger}
   ( \partial^{\mu} Y^A + i B^{\mu} Y^A  )
 - V_{\rm bos}
\end{eqnarray}
The equations of motion are
\begin{eqnarray}
A^{\mu}&:& \qquad F_B^{\mu\nu} = 0 \label{eom_A}\\
B^{\mu}&:& \qquad \frac{k}{8\pi} \varepsilon^{\mu\nu\rho} F_{\nu\rho}
 - 2 B^{\mu} Y_A^{\dagger} Y^A 
 + i Y_A^{\dagger} \partial_{\mu} Y^A 
 - i Y^A \partial_{\mu} Y_A^{\dagger} = 0 \label{eom_B} \\
(Y_A)^{\dagger} &:& \qquad \partial_{\mu} \partial^{\mu} Y^A 
+ i (\partial_{\mu} B^{\mu}) Y^A + B_{\mu} B^{\mu} Y^A = 0
\label{eom_Y}
\end{eqnarray}
{}From (\ref{eom_A}), we can put $B_{\mu}=0$ in some proper
gauge.\footnote{
If we introduce the scalar $a$ which is dual to $F_{\mu \nu}$,
taking this gauge means that taking $d a=0$ gauge.}
For simplicity, we put
$$
Y^A=0, \quad (A=2,3,4).$$
Now (\ref{eom_Y}) becomes the usual Laplace equation,
whose simplest solution is
\begin{eqnarray}
Y^1 = v \, e^{i p_\mu x^\mu}, \qquad p \cdot p = 0 
\label{pwave}
\end{eqnarray}
Substituting this and $B_{\mu}=0$ into (\ref{eom_B}),
we obtain
\begin{eqnarray}
 \frac{k}{16\pi}  \varepsilon^{\mu\nu\rho} F_{\nu\rho} 
= v^2 p^{\mu} ,
\label{solf}
\end{eqnarray}
which will be generalized 
to a linear combination of the solutions (\ref{pwave}) with
different $p_\mu$
by replacing the r.h.s. to the current of the $U(1)$,
which whose connection is $A_\mu$, of $Y^1$.
We find a constant flux is the solution of (\ref{solf}), 
therefore, (\ref{pwave}) with a constant flux is 
a solution of the ABJM action.
Note that because $p \cdot p=0$, 
the flux should be light like, which means that
if there is a non zero magnetic flux,
a nonzero electric flux also exists.

In the D2-brane limit $v \rightarrow \infty$ with $k/v$ fixed,
$F_{\mu \nu}$ becomes the field strength on 
the D2-brane and
the solution becomes a configuration in the D2-brane
with a constant light like flux and the dual photon $a$
is linear in $x^\mu$, namely $a \sim p_\mu x^\mu$.
Note that
the equations of motion for the dual photon in the D2-brane
is $*F \sim d a$.
In the D2-brane action, there are terms like
\begin{eqnarray}
S \sim \int F B + F \wedge C^{(1)} + \cdots, 
\end{eqnarray}
and we see that the $F^{0i}$ couple to the NSNS $B$-field
while $F^{12}$ couple to $C^{(1)}$.
A nonzero $F^{0i}$ flux correspond to infinite fundamental
strings smeared in D2-branes.
A nonzero $F^{12}$ correspond to
D0-branes smeared in D2-branes.
Lifting this situation into M-theory,
there exist charges of the KK-mode and M2-branes 
wrapping the M-circle
apart from the original M2-brane charge.
Combining the original M2-brane and wrapped M2-branes,
we obtain a helical M2-brane.
This interpretation is consistent with 
the M2-brane configuration (\ref{pwave}).

The discussion above is consistent with
the interpretation that "M0-brane"
is momentum for $S^1$ direction.

\section{Conclusion and Discussion}

In this paper we have constructed 
the classical solution in the ABJM theory
corresponding to M5-branes with a non-zero self-dual three-form flux,
from the solutions in the D2-brane action
in the scaling limit $k \rightarrow \infty$. 
We discussed that our solution is closely related with
the three-algebra.
We also found another solution, 
which correspond to an M2-brane winding the M-circle 
with momentum in the M-circle.

The brane charges are computed from the central charge 
of the supersymmetry algebra for the ABJM action in \cite{Low}.
Because our solution does not depend on the world volume coordinates 
of the M2-branes, a possibly non-vanishing central charge is 
${\cal Z}^{AB}_{EF}$. This may correspond to the D6-brane charge
as argued in the BLG model \cite{Furuuchi}
and, indeed, we confirm that this vanishes non-trivially 
by an explicit calculation.
The M5-brane charge would appear if we can include the 
non-linear supersymmetry transformation as done in the BLG case \cite{Furuuchi}.
However, the $\mathbb{Z}_k$ orbifolding eliminating it
and M5-brane charge does not appear in the central charges.
 
It would be interesting to generalize the construction of the 
solution to other branes.
It was shown that the D3-brane action with a nonzero 
$\theta$ term can be derived from 
the orbifolded ABJM action \cite{Fuji, Benna:2008zy, Imamura, TeYa} 
in an appropriate limit \cite{Hashimoto}.
It will be possible to find 
solutions in M2-brane from solutions in the D3-brane action
and other D-brane action.
For example, the instantons in the D3-brane action
will correspond to the M0-M2 system.

Near the orbifold singularity, the solution will be not valid,
then we need $r \gg l_p$.\footnote{
In order to match the tension, we used the large $B$ limit of the 
DBI action without the higher derivative corrections of the D4-brane.
This will be valid for $r \gg l_s$.}
Probably we need to resolve the singularity at the origin 
or introducing other branes for justifying the validity of the solution.
Furthermore, the solution is non-BPS, thus the stability and quantum
corrections to the solutions are also important. 
Most important thing to be studied for our solution
is the hidden $U(1)$ direction, which should be related 
to non-perturbative effects including monopoles.
We hope to investigate these problems in near future.

\section*{Acknowledgments}
We would like to thank K. Hosomichi, Y. Imamura, S. Sugimoto, and Piljin Yi for 
useful discussions.
The authors thank the Yukawa Institute for Theoretical Physics at Kyoto University. Discussions during the YITP workshop YITP-W-09-04 on ``Development of Quantum Field Theory and String Theory'' were useful to complete this work. 
S.~T.~is partly supported by the Japan Ministry of Education, Culture, Sports, Science and Technology. 

\appendix

\section{From D2-brane Action to 3D SYM Action}\label{DBI_SYM}

In this appendix, we briefly review the derivation
of 3D Super Yang-Mills (SYM) action from the DBI action of D2-brane
in order to fix the normalization of the gauge coupling constant
and of scalar fields.

For simplicity, we consider DBI action of a single D2-brane
in a flat background.
The action is 
\begin{eqnarray}
S_{\rm D2} = - T_{\rm D2} \int d^3 \sigma \sqrt{- {\rm det}(g+2\pi l_s{}^2 F)},
\label{DBI}
\end{eqnarray}
where $T_{\rm D2}$ is the D2-brane tension
\begin{eqnarray}
T_{\rm D2} = \frac{1}{( 2\pi )^2 g_s l_s{}^3} 
\left( = \frac{1}{(2\pi)^2 l_p{}^3} = T_{\rm M2} \right), 
\end{eqnarray}
and $g$ is the induced metric
\begin{eqnarray}
g_{\alpha\beta} = \partial_{\alpha} X^{\mu} \partial_{\beta} X^{\nu} \eta_{\mu\nu}. \qquad
\alpha,\beta = 0,1,2 , \qquad \mu,\nu = 0, \cdots , 9
\end{eqnarray}
By expanding this action in terms of the field strength $F$, 
we obtain 
\begin{eqnarray}
S_{\rm D2} = - T_{\rm D2} \int d^3 \sigma \sqrt{- {\rm det} g}
\left( {\bf 1} + \frac{(2 \pi l_s{}^2)^2}{4} F_{\alpha \beta} F^{\alpha \beta} 
+ {\cal O} (F^4) \right)
\end{eqnarray}
By comparing the second term with the gauge kinetic term of the supersymmetric U(1) gauge theory 
\begin{eqnarray}
- \frac{1}{4g_{\rm YM}{}^2} \int d^3 \sigma F_{\alpha\beta} F^{\alpha\beta},
\end{eqnarray}
we find 
\begin{eqnarray}
\frac{1}{g_{\rm YM}{}^2} = \frac{l_s}{g_s}.
\label{gYM_gs}
\end{eqnarray}

In order to derive the scalar kinetic term,
we need to use static gauge, in which 1+2 scalars are fixed as 
\begin{eqnarray}
X^0 = \sigma^0, \quad X^2 = \sigma^1, \quad X^3 = \sigma^2, 
\end{eqnarray}
while 7 scalars remain dynamical, which we relabel by the indices $i=1,\cdots 7$.
The scalar fields $X^i$ correspond to the transverse direction of the D2-brane in this gauge.
By expanding the volume factor in terms of scalar fields in this gauge, we obtain
\begin{eqnarray}
\sqrt{-g} 
&=& \sqrt{-{\rm det} (\eta_{\alpha\beta} 
+ \partial_{\alpha} X^i \partial_{\beta} X^i)} \nonumber \\
&=& 1 - \frac{1}{2} \partial^{\alpha} X^i \partial_{\alpha} X^i
+ {\cal O}(X^4)
\end{eqnarray}
Thus, the scalar kinetic term is given by
\begin{eqnarray}
- \frac{1}{2} \frac{1}{(2\pi)^2 l_p{}^3} \partial^{\alpha} X^i \partial_{\alpha} X^i
+ {\cal O}(X^4),
\end{eqnarray}
where we used $l_p$ instead of $l_s$ and $g_s$ for later convenience.
We should rescale the scalar fields as
\begin{eqnarray}
X^i = 2 \pi l_p{}^{3/2} \Phi^i
\label{XPhi}
\end{eqnarray}
in order to make the scalar kinetic term canonical.
The normalizations of the gauge kinetic term (\ref{gYM_gs})
and of the scalar fields (\ref{XPhi})
are also valid for multiple D2-branes.

\section{From ABJM Action to 3D SYM Action}

In this section, we review the reduction
from ABJM action to 3D SYM action
in the scaling limit.
Through this procedure, the relation (\ref{limit1}) is derived
including numerical factor.

The bosonic part of the ABJM action is given by
(\ref{ABJM_action}) and (\ref{Potential}).
When we put
\begin{eqnarray}
A^{(+)} \equiv \frac{1}{2} 
\left( A_{\mu}^{(1)} + A_{\mu}^{(2)} \right), \qquad
A^{(-)} \equiv \frac{1}{2} 
\left( A_{\mu}^{(1)} - A_{\mu}^{(2)} \right),
\end{eqnarray}
and
\begin{eqnarray}
D_{\mu} Y \equiv \partial _{\mu} Y + i [A^{(+)}_{\mu}, Y],
\end{eqnarray}
the Lagrangian is written as 
\begin{eqnarray}
L &=& \frac{k}{2\pi} \varepsilon^{\mu\nu\rho} 
{\rm tr} \left( A^{(-)}_{\mu} F^{(+)}_{\nu\rho} 
+ \frac{2i}{3} A^{(-)}_{\mu} A^{(-)}_{\nu} A^{(-)}_{\rho} \right) \nonumber \\
&&- {\rm tr} \left[ \left( D_{\mu} Y_A + i \{ A^{(-)}_{\mu} , Y_{A} \} \right)^{\dagger} 
\left( D^{\mu} Y^A + i \{ A^{(-)}_{\mu} , Y_{A} \} \right)  \right]
- V_{\rm bos},
\end{eqnarray}
where $V_{\rm bos}$ is the same as (\ref{Potential}).

Here, we suppose that one of the scalar fields obtain a vacuum expectation value
\begin{eqnarray}
\langle Y^4 \rangle = v {\bf 1}_{N \times N}.
\end{eqnarray}
Expanding the theory around this vacuum,
taking the scaling limit:
\begin{eqnarray}
v, k \to \infty, \qquad v/k: {\rm fixed}
\end{eqnarray}
and integrating out $A^{(-)}$, we obtain
\begin{eqnarray}
L = - \frac{k^2}{32 \pi^2 v^2} {\rm tr} 
\left[ \left( F_{\mu\nu}^{(+)} \right) ^2 \right]
- \frac{k}{4 \pi v} \varepsilon^{\mu\nu\rho} {\rm tr} \left[ (D_{\mu} \Phi^8) F^{(+)}_{\nu\rho} \right]
\nonumber \\
- \frac{1}{2} {\rm tr} \left[ (D_{\mu} \Phi^i) (D^{\mu} \Phi^i) \right]
- \frac{2 \pi ^2 v^2}{k^2} 
{\rm tr} \left[ \Phi^i , \Phi^j \right] ^2
\label{LYM}
\end{eqnarray}
where we put
\begin{eqnarray}
Y^1 = \frac{1}{\sqrt{2}} \left( \Phi^1 + i \Phi^2 \right) , \quad 
Y^2 = \frac{1}{\sqrt{2}} \left( \Phi^3 + i \Phi^4 \right) , \quad \nonumber \\
Y^3 = \frac{1}{\sqrt{2}} \left( \Phi^5 + i \Phi^6 \right) , \quad 
Y^4 = \frac{1}{\sqrt{2}} \left( \Phi^7 + i \Phi^8 \right) , \quad 
\label{YPhi}
\end{eqnarray}
and $i=1,\cdots ,7$.
The factor $1/\sqrt{2}$ is needed in order to reproduce the coefficient
$1/2$ of the kinetic term for the real scalar fields $\Phi^i$.
The second term of (\ref{LYM}) is a total derivative and can be ignored,
and thus, (\ref{LYM}) can be seen as the action of the 3D SYM.
From the first term of (\ref{LYM}), we see that 
the gauge coupling constant of the 3D SYM is written as
\begin{eqnarray}
\frac{1}{4 g_{\rm YM}^2} = \lim _{k,v \to \infty}
\frac{k^2}{32 \pi^2 v^2}
\label{gYM}
\end{eqnarray}
Thus, the relation (\ref{limit1}) is derived.

\section{Non-commutative Parameter and M-circle Radius}\label{NC_param}

The scalar fields $X^i$ in DBI action 
is related with scalar fields $\Phi^i$ in 3D SYM theory as in (\ref{XPhi}).
And this $\Phi^i$ is related with complex scalar fields $Y^a$ in ABJM action 
as in (\ref{YPhi}).
Thus, we have the normalization
\begin{eqnarray}
X^i + i X^{i+1} = 2 \sqrt{2} \pi l_p{}^{3/2} Y^a.
\label{scalar_factor}
\end{eqnarray}

The numerical factor of the relation (\ref{B_theta})
between the non-commutative parameter $\Theta$
and the expectation value of the $B$-field
can be explained from this normalization.
When we write the D4-brane solution of D2-brane action as
\begin{eqnarray}
[X^i, X^j] = i \Theta^{ij}_{D4}.
\label{D4_sol}
\end{eqnarray}
it is known that this non-commutative parameter $\Theta_{D4}$
is related to the background $B$ field in the D4-brane action as
\begin{eqnarray}
\Theta^{ij}_{D4} = \left( \frac{1}{B} \right)^{ij}.
\end{eqnarray}
Our M5-brane solution reduces to this D4-brane solution,
but we introduce the non-commutative parameter $\Theta$ as
\begin{eqnarray}
[Y^1, Y^2] = i \Theta.
\end{eqnarray}
The difference of the normalization factor (\ref{scalar_factor})
of scalar fields 
causes the difference between 
the normalization of our non-commutative parameter $\Theta$
and that of $\Theta^{ij}_{D4}$ in (\ref{D4_sol}) as
\begin{eqnarray}
\Theta_{D4} = 8 \pi^2 l_p{}^3 \Theta,
\end{eqnarray}
which explains the numerical factor in (\ref{B_theta}).

The numerical factor of the radius of the M-circle (\ref{Radius}) can also be explained
from the normalization (\ref{scalar_factor}) of the scalar fields. 
Since the vacuum expectation value of the scalar fields $X^i$ in DBI action
is the space-time length,
the length between the orbifold fixed point and the 
place at the M2 branes 
can be expressed in terms of vacuum expectation value $v$ of $Y^1$ as
\begin{eqnarray}
({\rm Length}) = 2 \sqrt{2} \pi l_p{}^{3/2} v.
\end{eqnarray}
Because the internal space which M2-branes are probing is ${\bf R}^8 / {\bf Z}^k$,
the length of the M-circle, which appears in a scaling limit,
is 
$$2 \pi (2\sqrt{2} \pi l_p{}^{3/2} v) / k, $$
which indicates that the radius of the M-circle is
\begin{eqnarray}
R = 2\sqrt{2} \pi l_p{}^{3/2} v / k
\quad \Leftrightarrow \quad R^2 = 8 \pi^2 l_p{}^3 v^2 / k^2.
\label{radius}
\end{eqnarray} 
reproducing the relation (\ref{Radius}).


\begin{thebibliography}{999}
\parskip=-2pt

\bibitem{Witten:1995ex}
  E.~Witten,
  Nucl.\ Phys.\  B {\bf 443}, 85 (1995)
  [arXiv:hep-th/9503124].


\bibitem{ABJM}
  O.~Aharony, O.~Bergman, D.~L.~Jafferis and J.~Maldacena,
  arXiv:0806.1218 [hep-th].

\bibitem{BL1}
  J.~Bagger and N.~Lambert,
  Phys.\ Rev.\  D {\bf 75}, 045020 (2007)
  [arXiv:hep-th/0611108].

\bibitem{BL2}
  J.~Bagger and N.~Lambert,
  Phys.\ Rev.\  D {\bf 77}, 065008 (2008)
  [arXiv:0711.0955 [hep-th]].

\bibitem{BL3}
  J.~Bagger and N.~Lambert,
  JHEP {\bf 0802}, 105 (2008)
  [arXiv:0712.3738 [hep-th]].

\bibitem{G}
  A.~Gustavsson,
  arXiv:0709.1260 [hep-th].

\bibitem{Terashima}
  S.~Terashima,
  arXiv:0807.0197 [hep-th].

\bibitem{Basu:2004ed}
  A.~Basu and J.~A.~Harvey,
  Nucl.\ Phys.\  B {\bf 713}, 136 (2005)
  [arXiv:hep-th/0412310].

\bibitem{GRVV}
  J.~Gomis, D.~Rodriguez-Gomez, M.~Van Raamsdonk and H.~Verlinde,
  JHEP {\bf 0809}, 113 (2008)
  [arXiv:0807.1074 [hep-th]].

\bibitem{HL}
  K.~Hanaki and H.~Lin,
  JHEP {\bf 0809}, 067 (2008)
  [arXiv:0807.2074 [hep-th]].
  

\bibitem{Fujimori}
  T.~Fujimori, K.~Iwasaki, Y.~Kobayashi and S.~Sasaki,
  JHEP {\bf 0812}, 023 (2008)
  [arXiv:0809.4778 [hep-th]].

\bibitem{Arai:2008kv}
  M.~Arai, C.~Montonen and S.~Sasaki,
  JHEP {\bf 0903}, 119 (2009)
  [arXiv:0812.4437 [hep-th]].

\bibitem{Kawai:2009rc}
  S.~Kawai and S.~Sasaki,
  Phys.\ Rev.\  D {\bf 80}, 025007 (2009)
  [arXiv:0903.3223 [hep-th]].

\bibitem{Kim:2009ny}
  C.~Kim, Y.~Kim, O.~K.~Kwon and H.~Nakajima,
  Phys.\ Rev.\  D {\bf 80}, 045013 (2009)
  [arXiv:0905.1759 [hep-th]].

\bibitem{Auzzi}
  R.~Auzzi and S.~Prem Kumar,
  arXiv:0906.2366 [hep-th].


\bibitem{Nastase:2009ny}
  H.~Nastase, C.~Papageorgakis and S.~Ramgoolam,
  JHEP {\bf 0905}, 123 (2009)
  [arXiv:0903.3966 [hep-th]].

\bibitem{HM}
  P.~M.~Ho and Y.~Matsuo,
  JHEP {\bf 0806}, 105 (2008)
  [arXiv:0804.3629 [hep-th]].

\bibitem{HIMS}
  P.~M.~Ho, Y.~Imamura, Y.~Matsuo and S.~Shiba,
  JHEP {\bf 0808}, 014 (2008)
  [arXiv:0805.2898 [hep-th]].





\bibitem{PSST}
  P.~Pasti, I.~Samsonov, D.~Sorokin and M.~Tonin,
  arXiv:0907.4596 [hep-th].


\bibitem{Low:2009de}
  A.~M.~Low,
  arXiv:0909.1941 [hep-th].


\bibitem{Krishnan:2008zm}
  C.~Krishnan and C.~Maccaferri,
  JHEP {\bf 0807}, 005 (2008)
  [arXiv:0805.3125 [hep-th]].


\bibitem{BL4}
  J.~Bagger and N.~Lambert,
  arXiv:0807.0163 [hep-th].
  
\bibitem{Benna:2008zy}
  M.~Benna, I.~Klebanov, T.~Klose and M.~Smedback,
  arXiv:0806.1519 [hep-th].



\bibitem{Mukhi}
  S.~Mukhi and C.~Papageorgakis,
  JHEP {\bf 0805}, 085 (2008)
  [arXiv:0803.3218 [hep-th]].

\bibitem{Perry}
  M.~Perry and J.~H.~Schwarz,
  Nucl.\ Phys.\  B {\bf 489}, 47 (1997)
  [arXiv:hep-th/9611065].

\bibitem{Schwarz}
  J.~H.~Schwarz,
  Phys.\ Lett.\  B {\bf 395}, 191 (1997)
  [arXiv:hep-th/9701008].

\bibitem{PST}
  P.~Pasti, D.~P.~Sorokin and M.~Tonin,
  Phys.\ Lett.\  B {\bf 398}, 41 (1997)
  [arXiv:hep-th/9701037].




\bibitem{Low}
  A.~M.~Low,
  JHEP {\bf 0904}, 105 (2009)
  [arXiv:0903.0988 [hep-th]].

\bibitem{Furuuchi}
  K.~Furuuchi, S.~Y.~Shih and T.~Takimi,
  JHEP {\bf 0808}, 072 (2008)
  [arXiv:0806.4044 [hep-th]].


\bibitem{Fuji}
  H.~Fuji, S.~Terashima and M.~Yamazaki,
  Nucl.\ Phys.\  B {\bf 810}, 354 (2009)
  [arXiv:0805.1997 [hep-th]].

\bibitem{Imamura}
  Y.~Imamura and K.~Kimura,
  Prog.\ Theor.\ Phys.\  {\bf 120}, 509 (2008)
  [arXiv:0806.3727 [hep-th]].

\bibitem{TeYa}
  S.~Terashima and F.~Yagi,
  JHEP {\bf 0812}, 041 (2008)
  [arXiv:0807.0368 [hep-th]].

\bibitem{Hashimoto}
  K.~Hashimoto, T.~S.~Tai and S.~Terashima,
  JHEP {\bf 0904}, 025 (2009)
  [arXiv:0809.2137 [hep-th]].






\end{thebibliography}
\end{document}